**Photonic Integrated Circuit for Rapidly Tunable Orbital Angular Momentum Generation Using Sb$_2$Se$_3$ Ultra-Low-Loss Phase Change Material**

*Md Shah Alam[1], Rudra Gnawali[2,*], Joshua R. Hendrickson[3], Diane Beamer[2], Tamara E. Payne[2], Andrew Volk[2], and Imad Agha[1,*]*

[1]Department of Electro-Optics and Photonics, University of Dayton, Dayton, OH 45469
[2]Applied Optimization, Inc. 3040 Presidential Dr. Suite 100 Fairborn, OH 45324
[3]Air Force Research Laboratory, Sensors Directorate, Wright-Patterson Air Force Base, OH 45433
iagha1@udayton.edu, rudra.gnawali@appliedo.com,

**Abstract**
The generation of rapidly tunable Optical Vortex (OV) beams is one of the most demanding research areas of the present era as they possess Orbital Angular Momentum (OAM) with additional degrees of freedom that can be exploited to enhance signal-carrying capacity by using mode division multiplexing and information encoding in optical communication. Particularly, rapidly tunable OAM devices at a fixed wavelength in the telecom band stir extensive interest among researchers for both classical and quantum applications. This article demonstrates the realistic design of a Si-integrated photonic device for rapidly tunable OAM wave generation at a 1550-nm wavelength by using an ultra-low-loss Phase Change Material (PCM) embedded with a Si-ring resonator with angular gratings. Different OAM modes are achieved by tuning the effective refractive index using rapid electrical switching of Sb$_2$Se$_3$ film from amorphous to crystalline states and vice versa. The generation of OAM waves relies on a traveling wave modulation of the refractive index of the micro-ring, which breaks the degeneracy of oppositely oriented whispering gallery modes. The proposed device is capable of producing rapidly tunable OV beams, carrying different OAM modes by using electrically controllable switching of ultra-low-loss PCM Sb$_2$Se$_3$.

**Introduction**
Orbital Angular Momentum (OAM)[1] is an additional degree of freedom of a photon, in addition to polarization and spin, which can be exploited in numerous applications such as optical communications[2,3], quantum optics[4], optical trapping[5], and lithography[6]. Since the development of OAM properties of light beams, researchers have aimed to investigate various switching techniques between different modes. As different OAM modes are used as bits in both quantum and classical OAM optical communications, the switching between OAM modes can dramatically enhance the data rate[7].

Circularly polarized light contains Spin Angular Momentum (SAM) of $\sigma\hbar$ ($\sigma = \pm 1$) per photon. However, in 1992, Allen et al. demonstrated that for the Optical Vortex (OV) light also possesses OAM[8] which is expressed in $l\hbar$ ($l = \pm 1, \pm 2, \pm 3, \ldots$) per photon. This provides a theoretically infinite number of channels that can be multiplexed with existing beam states since $l$ can take any value[1–7]. These states can be added to the existing beam states that use wavelength, amplitude and phase, time, and polarization of light to encode information. For the OV, the value of the phase's integral of the electric field around a path enclosing the vortex is equal to $l\hbar$, where $l$ is the topological charge and $\hbar$ is Planck's constant divided by $2\pi$. This OAM constitutes an additional degree of freedom[3,9] as the quantum characteristics $l\hbar$ do not interfere with each other.

The generation of OAM beams requires manipulation of the polarization and spatial mode of laser light. There are several methods to generate OAM beams which involve q-plates and phase retarders[10], nematic spatial light modulators[11], spatially varying retarders[12], interferometric methods[13], and integrated photonics such as Si-ring resonators with azimuthally varying grating elements[7]. Among these methods, integrated photonics could potentially provide more flexibility for tuning different modes at high speeds. As an example, for Si-ring resonators with azimuthally varying grating elements, a single Whispering Gallery Mode (WGM) is excited by a unidirectional wave from the bus waveguide, either clockwise or counterclockwise, and the resonators emit the OAM radiation vertically. Other integrated optics approaches for generating OAM radiation include an Archimedean spiral-shaped waveguide[14–16], a collection of subwavelength cavities designed for broadband OAM vertical radiation[17], and phased array nanoantennas[18].

Since the unique advantage of using optical OAM relies on properly accessing multiple OAM states, it is therefore desirable to implement devices that generate rapidly tunable OAM states of light, especially in an integrated form to afford large scale-up and low-cost. There are several prior demonstrations that have produced different OAM states on-chip by coupling to different ports on a passive chip[19], adjusting the operating wavelength[7], or by introducing heaters into Si as a mechanism for tuning[20]. However, while such devices have shown the principle of tunability, fast tuning of OAM modes would benefit from better mechanisms for tuning than was previously explored.

In this paper, a realistic and highly realizable model of a compact, integrated silicon photonic circuit composed of a Si-ring resonator with angular grating elements, embedded with ultra-low-loss $Sb_2Se_3$ Phase Change Material (PCM)[21] is presented to generate continuously and rapidly tunable OAM states. The compactness and flexibility of the device hold enormous promise for integration with other silicon photonic components for a wide range of applications.

**Integrated photonics for cylindrical vector vortices (CVVs)**

**Figure1 (a)** represents the schematic of the OAM generator and transceiver. This integrated photonic device has three basic elements: an access waveguide, circular resonator, and angular gratings on the inner wall of the circular resonator. When the access waveguide is excited by a laser beam, the light is coupled into the ring resonator. At resonance conditions, the the WGMs are produced and the angular grating elements emit a vortex beam carrying OAM travelling normal to ring resonator plane[7]. The emitted vortex beam are also referred to as cylindrical vector vortices (CVVs) as it exhibits cylindrically symmetric intensity distribution and polarization[22,23]. The resonance mode of the emitted waves depends on the number of grating elements, radius, effective refractive index of the ring resonator, and the wavelength. Usually, the WGMs are low loss which makes it very difficult to get light into or out of these modes. However, the angular gratings inside the ring resonators provide a periodic variation of the dielectric constant which helps in coupling the unperturbed Eigen-modes of the resonator into other modes. The gratings scatter the WGMs as well-controlled OAM states for free-space propagation in a similar fashion to waveguide grating couplers[24] scattering Gaussian modes, acting as an optical vortex emitter. When the quasi-Transverse Electric (TE) mode of different laser wavelengths are applied into the access waveguide, various cavity resonances are excited, and azimuthally polarized vector vortices with quantized OAM waves are emitted from the device. Conversely, when an azimuthally polarized vortex beam of different OAM state is applied towards the ring resonator containing the grating elements, then at the resonance conditions, the WGMs are confined into the ring resonator and back-coupled into the access waveguide so that the device acts as an OAM detector[24].

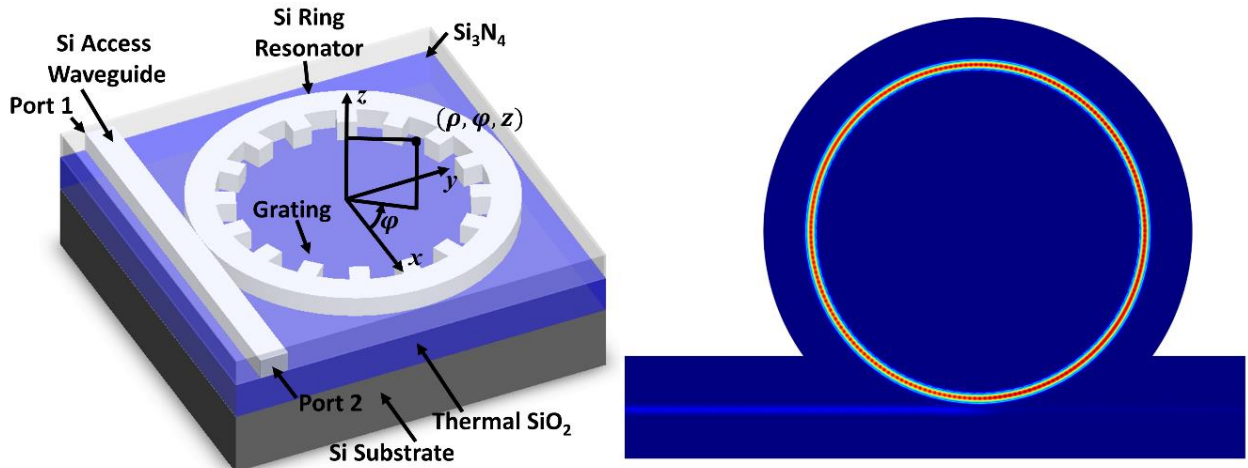

**Figure 1.** (a) Schematic of the Si-ring resonator with angular grating elements for OAM generation. (b) Electric field intensity inside the ring and access waveguide at resonance wavelength 1550.85 nm.

The Jones vector of the output CVVs carrying OAM can be represented by[25,26]:

$$\boldsymbol{E}_{out} \propto \frac{1}{2}\left\{\sqrt{1+\sigma}e^{i(l_{TC}+1)\varphi}\begin{bmatrix}1\\i\end{bmatrix} + \sqrt{1-\sigma}e^{i(l_{TC}-1)\varphi}\begin{bmatrix}1\\-i\end{bmatrix}\right\} \quad (1)$$

Where, $l_{TC} = p - q$ is the Topological Charge (TC)[7], and $\sigma(|\sigma| \leq 1)$ is the transverse spin state in the near field evanescent wave at the sidewalls of the ring resonator due to grating perturbation. For the transverse spin angular momentum (SAM) of left (right) handed spin $\sigma > 0$ ($< 0$). $q$ and $p$ are the numbers of the grating elements around the circular resonator and the WGM azimuthal mode order (numbers of the optical periods in WGMs), respectively. The number of the optical periods is defined as:

$$p = \frac{2\pi R n_{eff}}{\lambda} \quad (2)$$

where $n_{eff}$ and $R$ are the effective index of the WGMs and the inner radius of the circular resonator, respectively, and $\lambda$ is the wavelength of illumination which is around 1550 nm for the proposed design.

The total angular momentum of the emitted paraxial optical vortex beam can be considered as the sum of the OAM and SAM components which are associated with the spatial properties and polarization of light respectively[26]. The angular phase variation in the transverse component of CVVs is a linear function of the cylindrical coordinate $\varphi$ and can be written as[26]:

$$\Phi_p = (l_{TC}\varphi \pm \sigma\varphi) \quad (3)$$

Where, $l_{TC}\varphi = (p - q)\varphi$ arises from the first order diffraction of the grating elements (OAM) and $\pm\sigma\varphi$ corresponds to local transverse spin state (SAM) while counter clockwise CCW(+) or clockwise CW(−) WGMs travel around the ring resonator. When the polarization at the grating scatterer location reaches one of the circular polarizations (LHCP: $\sigma = +1$, RHCP: $\sigma = -1$) then the superposition reduces to $(l_{TC} \pm 1)\varphi$. On the other hand, the z component (along the CVVs propagation direction) of the SAM and OAM carried by the CVVs can be represented by[26]:

$$S_z = \sigma\hbar, \quad L_z = (l_{TC} - \sigma)\hbar \quad (4)$$

where, $\hbar$ is the reduced plunk constant. Here, the Total Angular Momentum (TAM) of the z component,

$$J_z = S_z + L_z = l_{TC}\hbar = (p - q)\hbar \quad (5)$$

is conserved with the azimuthal mode order $p$ of WGM and number of grating elements $q$ regardless of the transverse spin state (SAM). This is due to the rotationally symmetric anisotropy orientation of the scatterer grating elements.

Here, an integrated OAM generator and transceiver is designed using a Si ring resonator which is built on top of standard 220-nm-thick Si-on-Insulator (SOI) wafer. The width of the access waveguide and ring resonator is 550 nm, and the inner radius of the ring resonator is 9.45 μm. The length and width of the grating elements are kept at 15% of the grating period (Λ). The gap between the access waveguide and the circular resonator is kept at 120 nm. The number of grating elements is chosen to be 104 in order to achieve an OAM wave with a topological charge $(l_{TC} = p - q) = 0$ within the range of 1540 nm to 1550 nm. The Si waveguide and the ring resonator with grating elements are on top of a 2-μm thermal oxide ($SiO_2$) and buried in 1.5 μm of LPCVD/PECVD Si3N4. The topological charge $(l_{TC} = p - q)$ can take any integer value depending on the difference of $p$ and $q$. According to **Equation 2**, p increases with the increase of the effective refractive index of the ring resonator but decreases with the increase of the wavelength of the excited beam for a fixed radius ring resonator. A three dimensional finite element method model (COMSOL) is used to perform numerical simulation to find the resonance wavelengths for different OAM modes. Port 1 (**Figure 1a**) is excited with TE 01 mode and wavelengths are tuned from 1510 nm to 1585 nm. **Figure 1b** shows the electric field intensity inside the ring resonator while it is at resonance and emits CVVs carrying OAM wave with $l = -1$. **Figure 2** represents the intensity and corresponding phase of the different emitted OAM waves radiated normal to the ring resonator plane towards $(+z)$ direction at different resonant wavelengths. According to **Equation 5** the z component of radiated CVVs represents total angular momentum which equals the topological charge of the OAM wave ($l_{TC}$). **Figure 2a** and **2b** represent the $|E_z|^2$ and the phase of the z component of the radiated wave respectively. According to **Figure 2b**, the proposed ring resonator radiates OAM waves with $l_{TC} = +3, +2, +1, 0, -1, -2, -3, -4$ at 1511.60, 1521.20, 1531.00, 1540.50, 1550.85, 1561.25, 1571.55, and 1581.90 respectively. The total angular momentum $l_{TC} \pm 1$ (due to OAM and SAM in the transverse components ($I_x, I_y$)) depends the transverse spin $\sigma$ (+/-) and rotation of the WGMs (CCW/CW) in the ring resonator as can be observed in **Figure 2c** ($I$), **Figure 2d** (phase of $E_y$, $\varphi_y$), **Figure 2e** (phase of $E_x$, $\varphi_x$). **Figure 2f** represents the phase difference of $E_y$ and $E_x$, $\varphi_y - \varphi_y = \mp \pi/2$ which demonstrates the existence of SAM in CVVs along with the OAM. From **Figure 2d, 2e** and **2f**, it is observed that the radiated CVVs possess RHCP for $l_{TC} > 0$, while LHCP for $l_{TC} < 0$ along with the OAM. The Free Spectral Range (FSR) is found to be close to 10 nm. Increasing the radius of the ring resonator reduces the value of the FSR and increases the number of modes in a given spectral range.

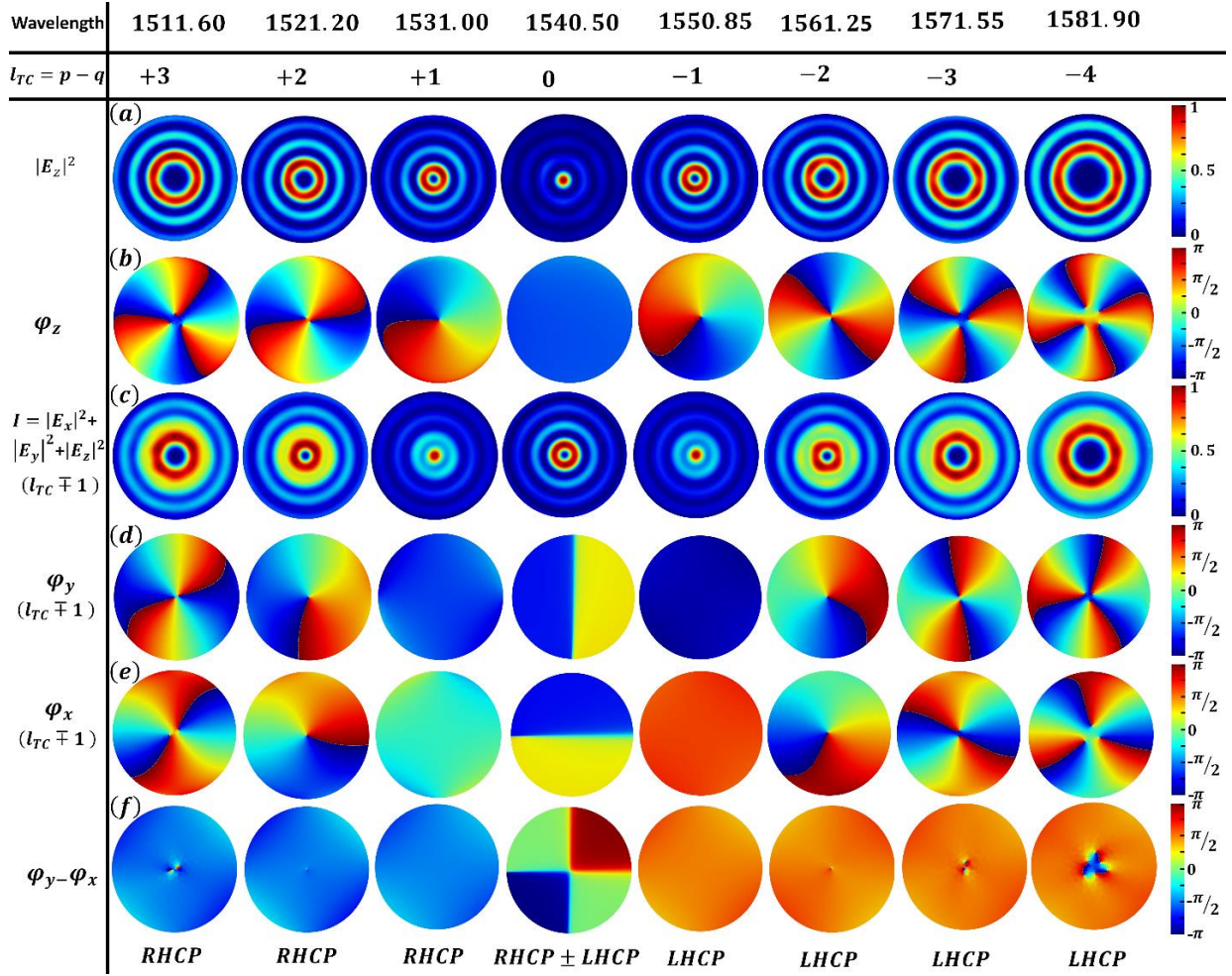

**Figure 2.** Generation of different OAM modes from Si ring resonator with grating elements by tuning wavelengths from 1510 to 1585 nm. Ring radius is 9.45 µm. Different OAM modes of $l = +3, +2, +1, 0, -1, -2, -3, -4$ is generated at 1511.60, 1521.2, 1531.00, 1540.50, 1550.85, 1561.25, 1571.55, 15081.90 respectively. **(a)** Longitudinal component of electric field intensity ($|E_z|^2$) of different OAM modes $l = +3, +2, +1, 0, -1, -2, -3, -4$ **(b)** Spiral phase of longitudinal component of electric field $E_z$ of different OAM modes. **(c)** Total electric field intensity of emitted CVVs containing total angular momentum (TAM) of $l_{TC} \pm 1$. **(d)** Phase of transverse $E_x$ of the emitted vortex beam (with topological charge $l_{TC} \pm 1$). **(e)** Phase of transverse $E_y$ component. **(f)** Phase difference between transverse $E_y$ and $E_x$ ($\mp \pi/2$) showing the presence of spin angular momentum (SAM) in the emitted vortex beam along with the OAM.

**Rapidly Tunable OAM generator and transceiver**

For a specific design, the number of grating elements is fixed. Therefore, tuning different OAM modes at a specific wavelength requires the change of the effective refractive index of the ring resonator hence changing the WGM order. Similarly, a large refractive index change is required to tune different modes for a larger FSR. At the expense of increasing the Si-ring size, the adjacent mode can be brought closer. However, to make the device more compact, we have considered a 9.45-µm ring in which the adjacent modes are separated by approximately 10 nm.

Traditionally, PCMs allow for tuning of the refractive index due to the inherent dielectric constant contrast as the material transitions from a glassy to a crystalline state in response to a thermal stimulus[21,27–29]. As such, the effective refractive index of the ring resonator can be tuned by introducing a PCM film with a larger refractive index contrast between the amorphous and crystalline states. The PCM film can be closely placed on top of the ring resonator in the cladding region as shown in schematic diagram of **Figure 3a**. By recurrently switching the phase of the PCM, the effective refractive index can be rapidly varied to tune different OAM modes. There are several types of PCMs that can provide a significant amount of refractive index change while changing the phase from amorphous to crystalline states and vice versa. However, most of them are very lossy at the telecom band. For example, a complex refractive indices for amorphous and crystalline states are $4.2402 + i\,0.05094$ and $6.1095 + i\,0.72127$, respectively, for GST[30] and $4.2 + i\,0.034$ and $5.5 + i0.27$, respectively, for GeTe[31]. Therefore, they are not suitable for designing a tunable

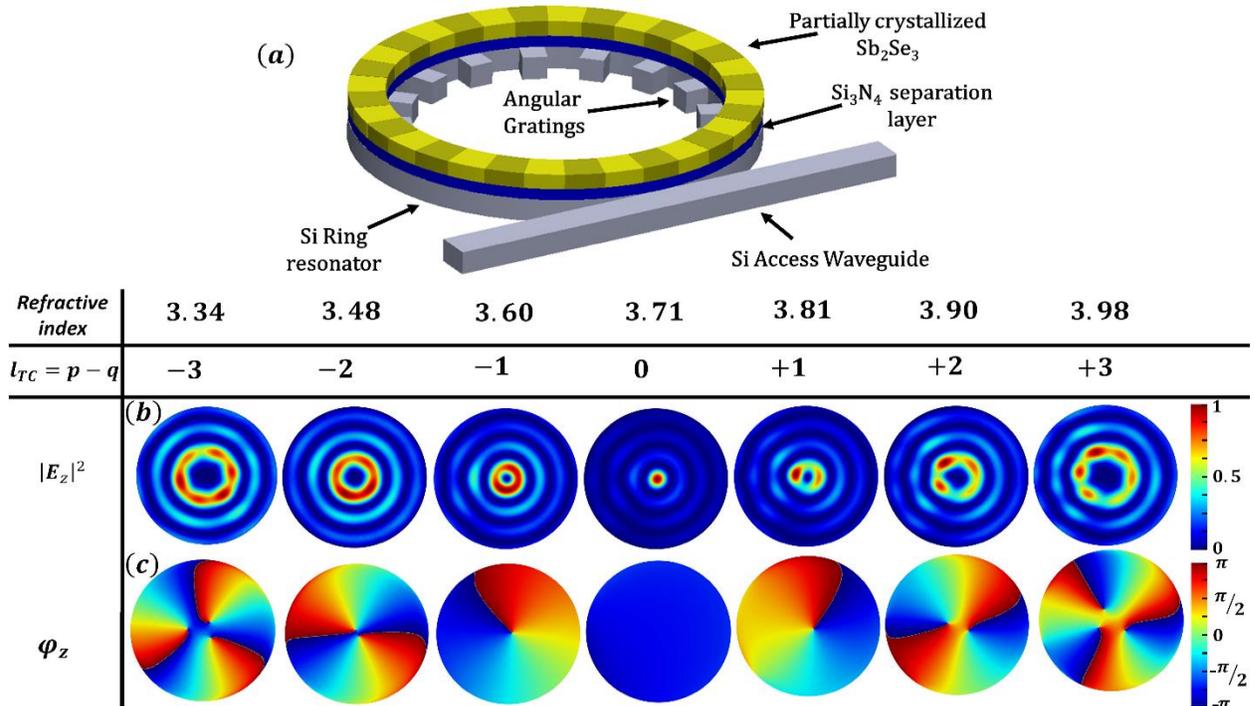

**Figure 3.** Generation of different modes of OAM waves at 1550 nm by embedding Sb$_2$Se$_3$ PCM: (a) Schematic of tunable OAM generator consisting of Si waveguide, ring resonator with angular gratings, and 105-nm-thick Sb$_2$Se$_3$ PCM placed on top of the ring with a separation of 50 nm. The Si waveguide is on top of a thermally evaporated SiO$_2$ layer and buried by a Si$_3$N$_4$ layer. **(b)** Longitudinal component of electric field intensity ($|\mathbf{E_z}|^2$) of different OAM modes $\mathbf{l = -3, -2, -1, 0, +1, +2, +3}$ **(c)** Spiral phase of longitudinal component of electric field E$_z$ of different OAM modes.

OAM device as any high-Q resonance will be washed out. However, Sb$_2$Se$_3$ and Sb$_2$S are promising ultra-low-loss PCMs at 1550 nm[21]. The amorphous and crystalline refractive indices of Sb$_2$Se$_3$ are $3.285 + i0$,

$4.05 + i0$ ($\Delta n = 0.765 + i0$) respectively and for the $Sb_2S$ are $2.712 + i0$, $3.308 + i0$ ($\Delta n = 0.596 + i0$) respectively. Although they have a comparatively low refractive index contrast, but a thin film of $Sb_2Se_3$/$Sb_2S$ can be placed very close to the Si-ring resonator as they are lossless. Therefore, by switching the film between its two phases, a reasonable amount of effective refractive index change is incurred and thereby different OAM modes can be tuned. This can be attributed to the Figure of merit that defines optical PCMs: Mainly the contrast of refractive index vs the increase in extinction coefficient.

For the initial design, we have considered 110 grating elements in the 9.45 µm ring resonator and a 105-nm-thin $Sb_2Se_3$ film on top of the Si ring separated by 50 nm thin $Si_3N_4$ film to enable tuning between different OAM modes at 1550 nm wavelength. **Figure 3a** shows a schematic of the tunable OAM generation by partial phase change. **Figure 3b-c** represent the generation of different OAM modes by partially changing the phase of the $Sb_2Se_3$ film. Here, the refractive index of the $Sb_2Se_3$ film is considered to be between 3.285 (fully amorphous) and 4.05 (fully crystallized) as an effect of partial crystallization. From simulation, it is observed that $l_{TC} = -3$, $l_{TC} = -2$, $l_{TC} = -1$, $l_{TC} = 0$, $l_{TC} = +1$, $l_{TC} = +2$, $l_{TC} = +3$ can be achieved by partially changing the phase of the $Sb_2Se_3$ from amorphous to crystalline by considering overall refractive index as 3.34, 3.48, 3.60, 3.71, 3.81, 3.90 and 3.98 respectively.

**Comparison of different design options for tunability**

To achieve the desired fraction of phase change of the Sb2Se3 film, four different types of realistic designs using Indium Tin Oxide (ITO)-based nano-heater circuits are proposed as shown in **Figure 4a-d.** A rapid switching of Sb2Se3 film using joule heating of the ITO based nano-heater circuits can be achieved using the proposed design. The $Sb_2Se_3$ film needs to be cooled from above its crystallization temperature (200 °C) to crystallize the amorphous $Sb_2Se_3$ film[21]. To achieve reamorphization, the temperature should be raised to the melting point (610 °C) and needs to be cooled down faster than the crystallization speed, leaving it in an amorphous state.

The feasibility of the reversible switching for the proposed designs to tune OAM modes from $l = -3$ to $l = +2$ at 1550 nm depends on the effective change of the refractive index of the ring resonator which can be evaluated by the **Equation 6**:

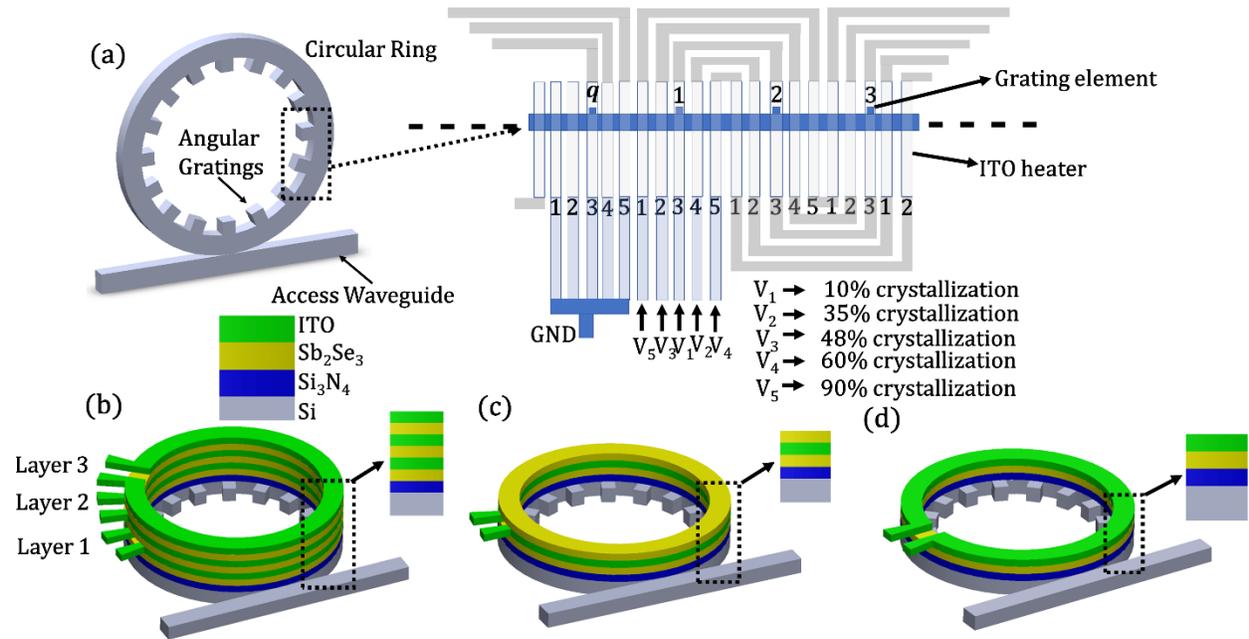

**Figure 4.** Schematic of ITO micro and nano heater-based OAM generator: **(a)** Design 1, **(b)** Design 2, **(c)** Design 3, and **(d)** Design 4

$$l = p - q = \frac{2\pi R n_{eff}}{\lambda} - q. \qquad (6)$$

For our design, the number of grating elements is 110, the inner ring radius is $R = 9.45$ μm, and $\lambda = 1550$ nm. Therefore, the effective index of the ring resonator should be tuned to 2.7932, 2.8193, 2.8454, 2.8715, 2.8976, and 2.9237 to generate $l = -3, -2, -1, 0, +1, +2$, respectively. The phase of the appropriate fraction of cladding $Sb_2Se_3$ film needs to be changed by applying a pulse voltage of proper magnitude and pulse duration on the nano-heater to achieve the required effective refractive index to tune the OAM waves from $l = -3$ to $l = +2$.

**Design 1: Figure 4a** represents the schematic of an ITO nano-heater based OAM generator. A 220-nm-thick Si-ring resonator with a 9.45 μm inner radius and 110 azimuthal grating elements and a closely placed access waveguide can be used to generate OAM waves. The ring resonator and access waveguide are covered with a very thin layer of $Si_3N_4$. A $Sb_2Se_3$ PCM ring of the same radius as the Si ring is placed on top of the Si-ring resonator. Five different ITO nano-heaters are placed in between each grating element as shown schematically in **Figure 4a**. The nano-heaters are connected in a series with the corresponding nano-heaters of each grating element according to the connection shown in **Figure 4a**. Each ITO nano-heater is connected to a different ultra-short pulse voltage source. $V_1$, $V_2$, $V_3$, $V_4$, and $V_5$ are applied at Port 3, Port 4, Port 2, Port 5, and Port 1 respectively corresponding to the 1st grating element. All five ports of the 110th grating element are connected to the electrical ground. Initially, the high fraction of the $Sb_2Se_3$ PCM is considered to be in an amorphous state and excites an OAM wave of $l = -3$. By applying a pulse voltage, $V_1$, at Port 3, a small fraction of the $Sb_2Se_3$ element is crystallized to excite an OAM wave of the $l = -2$ mode. Similarly, by applying $V_1$ and $V_2$, an $l = -1$ mode, $V_1$, $V_2$ and $V_3$ an $l = 0$ mode, $V_1$, $V_2$, $V_3$ and $V_4$ an $l = +1$ mode and $V_1$, $V_2$, $V_3$, $V_4$ and $V_5$ an $l = +2$ can be excited. The amorphization voltage, a comparatively higher voltage with a shorter pulse, is then applied to all of the ITO nano-heaters to reamorphize the necessary fraction of the $Sb_2Se_3$ ring to achieve an $l = -3$ mode. Initially, the required pulse voltage amplitude and pulse duration can be experimentally evaluated.

**Table 1.** Generation of different OAM modes by applying pulse voltages at different Layers.

| | | | |
|---|---|---|---|
| - | - | - | $-3$ |
| $L_3$ | | - | $-2$ |
| | $L_2$ | | $-1$ |
| $L_3$ | $L_2$ | | $0$ |
| | $L_2$ | $L_1$ | $+1$ |
| $L_3$ | $L_2$ | $L_1$ | $+2$ |
| $L_3$ (higher voltage shorter pulse) | $L_2$ (higher voltage shorter pulse) | $L_1$ (higher voltage shorter pulse) | $-3$ |

**Design 2: Figure 4b** represents a tunable OAM wave generator with multilayer $Sb_2Se_3$ PCM and ITO nano-heaters. In this case, three different layers of ITO rings are placed on top of three different $Sb_2Se_3$ rings of the same radius as the Si ring instead of five segmented ITO nano-heaters. Similar to design 1, the Si ring resonator with azimuthal grating elements and an access waveguide is buried within a very thin (50 nm) $Si_3N_4$ layer. Then, three successive thin $Sb_2Se_3$ ring layers and ITO ring layers are placed on top of the Si ring resonator. Layers 1 (**L₁**), 2 (**L₂**), and 3 (**L₃**) of ITO rings are connected to three different ultra-short pulse voltage sources $V_1$, $V_2$, and $V_3$, respectively. The layer closest to the Si ring resonator (**L₁**) has more impact while **L₃** has lowest impact in affecting the overall effective refractive index of the Si ring resonator by changing the phase of the PCM film. Initially high fraction of Sb₂Se₃ layers are considered to be in an amorphous state which can excite $l = -3$. Therefore, by applying different voltages at different layers other OAM modes can be excited according to **Table 1**. After achieving an $l = +2$ OAM mode, higher voltages with a shorter pulse are applied on all nano-heater layers (**L₁**, **L₂**, and **L₃**), and the appropriate fraction of the Sb₂Se₃ rings can be brought back to an amorphous state to achieve an $l = -3$ OAM mode.

**Design 3: Figure 4c** represents another tunable OAM generator consisting of bi-layer Sb₂Se₃ rings and only one ITO ring nano-heater sandwiched between them. This is a more simplified design compared to designs 1 and 2. Similar to designs 1 and 2, the Si-ring resonator and access waveguide are buried within a very thin Si₃N₄ layer, and two thin Sb₂Se₃ ring layers are separated by a thin ITO nano-heater ring placed on top of it. A tunable OAM wave can be generated by applying a pulse voltage of different magnitudes or multiple pulse duration times. In this case, only one nano-heater acts as a heat source and raises the temperature of both the Sb₂Se₃ rings simultaneously. The magnitude of the pulse voltage and the pulse duration can be experimentally determined to excite different OAM modes.

**Design 4: Figure 4d** represents the simplest design for a tunable OAM generator. In this design, only one Sb₂Se₃ ring and one ITO nano-heater are used. By applying varying pulse voltages of different magnitudes or pulse duration, several OAM modes can be tuned by changing the phase of the appropriate fraction of the Sb₂Se₃ film ring from amorphous to crystalline and vice versa. In order to achieve same number of OAM modes as design 3, a comparatively thicker Sb₂Se₃ ring is required which needs to be reversibly switchable which can be challenging due to the poor thermal conductivity of the Sb₂Se₃ film.

**Design comparison summary:**
- The main advantage of the first design is that it provides more controllability by applying five different voltage pulses at the same layer. However, this design has two main challenges:
    1. Sb₂Se₃ film rings should be thicker to cover a specific number of OAM modes which needs to be reversibly switchable through thermal heating.
    2. The long, thin ITO nano-heaters passing through all grating elements are prone to break due to fabrication failure and overheating.
- The second design overcomes the drawbacks of the first design:
    1. It requires comparatively thinner Sb₂Se₃ films to accommodate more OAM modes.
    2. It uses three wider ITO rings as nano-heaters which are more robust and not prone to failure.
  
  However, the main disadvantages of this design are:
    1. It requires a multistep fabrication steps.
    2. Multiple ITO layers reduce the OAM wave intensity.
    3. The third Sb₂Se₃ film layer has less effect as it is comparatively far away from the Si-ring resonator.
- The third design is a simplified version of design 2. Therefore, it has the same advantages of design 2 but overcomes the drawbacks:
    1. It only requires three different layers (two Sb₂Se₃ layers and one ITO layer sandwiched in between) which reduces the fabrication complexity, and the OAM wave intensity is less affected.
    2. It requires only a single voltage source. In this design, different OAM modes can be achieved by tuning the magnitude and pulse width of the single voltage source. Although it uses few multilayers, it is a compromise between designs 1 and 2 with more tunability and a robust design with more outcomes.
- The fourth design is the simplest design and uses only a single layer of Sb₂Se₃ film and one ITO nano-heater. This design is easier to fabricate and tune to different OAM modes. However, the main challenge is that

1. it requires a thicker $Sb_2Se_3$ film compared to the design 3 for achieving same number of OAM modes which is difficult to fully reamorphize as the amorphization requires the $Sb_2Se_3$ film to melt and cool down at a higher rate than the crystallization speed. Therefore, it has the challenges to tune more OAM modes by reversible switching of comparatively thicker $Sb_2Se_3$ film compared to other designs which use comparatively thinner $Sb_2Se_3$ films that are split.

**Thermal modeling of the switching behavior of the $Sb_2Se_3$ Film using an ITO Nano-heater**

We have investigated the effectivity of the simplest design (design 4) with only one $Sb_2Se_3$ ring and one ITO nano-heater ring using rigorous finite element calculations. We have considered a 220-nm-thick, 550-nm-wide Si access waveguide and ring resonator. The device modeled is based on a standard SOI wafer.

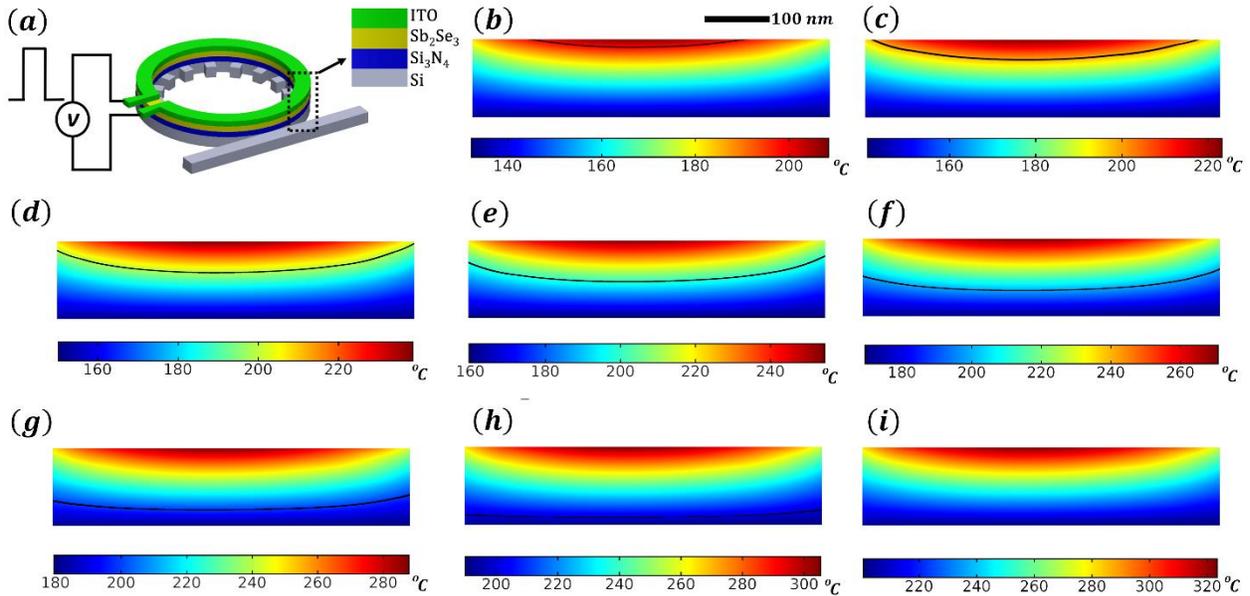

**Figure 5.** Temperature distribution in 105 nm $Sb_2Se_3$ film of tunable OAM generator while applying 1 μsec pulse voltage of different amplitude across the ITO nano-heater. (a) Schematic of the tunable OAM generator using single layer $Sb_2Se_3$ film and ITO film nano-heater ring. (b-i) temperature distribution along the cross-section of the $Sb_2Se_3$ film, the horizontal color bars represent the temperature in °C, the black contour line represent the boundary of 200 °C inside the $Sb_2Se_3$ film. (b) 13.0 volts, (c) 13.5 volts, (d) 14.0 volts, (e) 14.5 volts, (f) 15.0 volts, (g) 15.5 volts, (h) 16 volts, and (i) 16.5 volts.

Therefore, the Si waveguide and the ring resonator are on top of the thermal oxide ($SiO_2$) and buried in 50-nm Low Pressure Chemical Vapor Deposition (LPCVD) $Si_3N_4$. The 105-nm-thick $Sb_2Se_3$ ring, followed by the 120-nm-thick ITO nano-heater open ring with extended electrodes, are placed on top of the Si-ring resonator. The width of the $Sb_2Se_3$ ring and ITO nano-heater are considered 600 nm (slightly larger than that of the Si-ring resonator). The $Sb_2Se_3$ and ITO rings are covered by 1.5-μm-thick $Si_3N_4$ that could be grown via Plama Enhanced Chemical Vapor Deposition (PECVD) at a temperature lower than the melting temperature of $Sb_2Se_3$ (~610° C). We performed FEM simulations in COMSOL Multiphysics software to find the relation between the amplitude and pulse duration of the applied pulse voltage with the phase change of the fractional volume of the $Sb_2Se_3$ layer from amorphous to crystalline state. The material properties of $Sb_2Se_3$, ITO, $Si_3N_4$ and Si used in this simulation are taken from published literature[30,32–37]. **Figure 5** represents the cross-sectional view of the temperature distribution in $Sb_2Se_3$ film at 1 μsec when 1 μsec pulse voltage of different magnitude are applied. From **Figure 5(b-i)** it is observed that changing the magnitude of the pulse voltage from 13 volts to 16.5 volts the 105 nm Sb2Se3 can be transformed from partially to fully crystalize. Shown in **Figure 5**, around 3.5% of the volume fraction of the $Sb_2Se_3$ ring film exceeds the crystallization temperature of 200 °C for a 13-volt 1 μsec pulse. Around 15%, 30%, 45%,

60%, 80%, 90%, and 100% of the Sb$_2$Se$_3$ film reach 200 °C or above when 13.5, 14, 14.5, 15, 15.5, 16 and 16.5 volts of 1 μsec pulse respectively. Therefore, different volume fractions of the Sb$_2$Se$_3$ layer can be crystallized by applying 1 μsec pulse different magnitudes. This change of phase of Sb$_2$Se$_3$ introduces the change of effective refractive index of the Si-ring resonator with azimuthal grating elements, and different OAM modes can be generated while the access waveguide is excited with TE$_{01}$ mode of same wavelength. However, the most challenging aspect of this process is to reamorphize the crystallized Sb$_2$Se$_3$ layer because it requires the temperature to be raised above the melting point (610 °C) and to cool down faster than the crystallization kinetics. At the same time, the temperature of the ITO film must stay well below its melting point. The thermal conductivities of the PCM and the surrounding material are important for achieving the temperature of most or all of the volume fraction of the PCM film above the melting point while keeping the temperature of the ITO well below its melting point[30,38]. The thermal conductivity of ITO [34,39] and silicon nitride[35] are more than two order of magnitude higher than the thermal conductivity of Sb$_2$Se$_3$ (0.037 W/mK)[33].

**Figure 6** shows a comparative study of the temperature distribution and cooling rate inside the two different thick (105 nm corresponding to design 4 and bi-layer 85 nm corresponding to design 3) Sb$_2$Se$_3$ film when comparatively higher voltages of shorter pulse are applied in ITO nano-heater to almost completely melt the Sb$_2$Se$_3$ and bring the film back to the amorphous state by rapid cooling. The blue solid lines in **Figure 6** (**a** and **d**) represent the maximum temperature of ITO film of around 1250 °C and 1480 °C for the proposed designed (4 and 3) when 33 volts and 30 volts 650 ns pulse are applied across the ITO ring respectively. At this case, the Sb$_2$Se$_3$ film up to around 100 nm depth out of 105 nm depth is completely melted (**Figure 6 a**) >610 °C at 650 ns, and similarly up to around 75 nm depth out of 85 nm depth is melted (**Figure 6 d**) as shown by the purple lines. **Figure 6 b,e** represent temperature and cooling rate for the first 100 ns after the applied pulse voltage is switched off (from 650 ns to 750 ns). It is observed that within less than the 100 ns both the films (design 4: 105 nm Sb$_2$Se$_3$ and design 3: bi-layer 85 nm Sb$_2$Se$_3$) cool down to solid (< 610 °C). The cooling rate of film depends upon the temperature and thermal conductivity of the film and the surrounding materials. The cooling rate of ITO is initially much higher than that of $Sb_2Se_3$ as shown by the dotted blue lines and the thicker 120 nm ITO (**Figure 6b**) cools faster than thinner 85 nm ITO (**Figure 6e**). However, the thinner $Sb_2Se_3$ cools faster than that of the thicker $Sb_2Se_3$ as observed comparing the red, green, cyan and purple dotted lines in **Figure 6b** and **e**. From **Figure 6c** and **f** it is observed the 85 nm $Sb_2Se_3$ film (of proposed design 3) completely goes blow the crystallization temperature (200 $^oC$) within 1900 ns while it takes up to around 3000 ns for the 105 nm film (design 4) to reach below the crystallization temperature. Therefore, although it is possible to crystallize a thicker layer of $Sb_2Se_3$ film, but it may not be possible to reamorphize the most fraction of the comparatively thicker $Sb_2Se_3$ film.

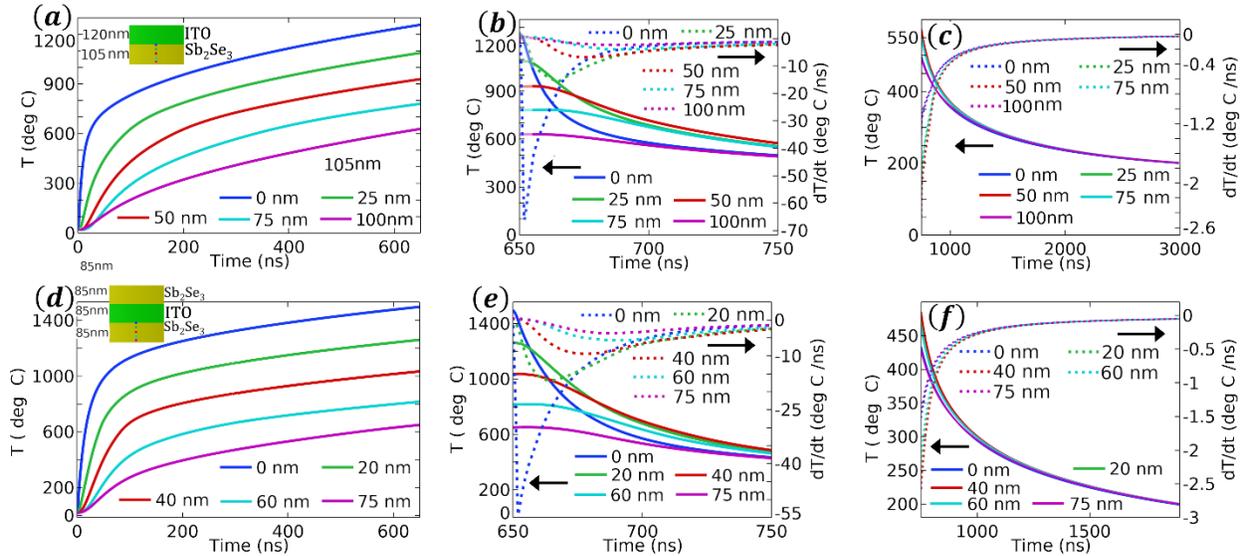

**Figure 6.** Temperature and cooling rate vs time at the center and along the depth of the ring (as shown in the inset of **Figure a** and **d**) for applied 650 ns pulse voltage across the ITO ring of tunable OAM generators. The solid lines represent temperature in (°C), dotted lines represent cooling rate. (a-c) corresponds to design 4: single layer 105 nm thick Sb2Se3 and 120 nm thick ITO ring; (d-f) correspond to design 3: bilayer 85 nm Sb$_2$Se$_3$ and sandwiched 85 nm thick ITO film ring). 33 volts and 30 volts are applied for design 4 and design 3 respectively. Blue lines (solid/dotted) represent the temperature and cooling rate of the ITO layer at the interface of ITO and Sb2Se3 film. The green, red, cyan and purple (solid/dotted) lines in (a-c) represent the temperature and cooling rate of Sb2Se3 film at 25 nm, 50 nm, 75 nm and 100 nm depth from the ITO and Sb2Se3 film layers' interface respectively. But, for (d-e) the green, red, cyan and purple (solid/dotted) lines correspond to that of 20 nm, 40 nm, 60 nm and 75 nm depth.

Therefore, to maintain the crystallization and reamorphization of a larger fraction of the $Sb_2Se_3$ film while maintaining tunability of more OAM modes, it is preferable to use design 3 which has two split 85-nm

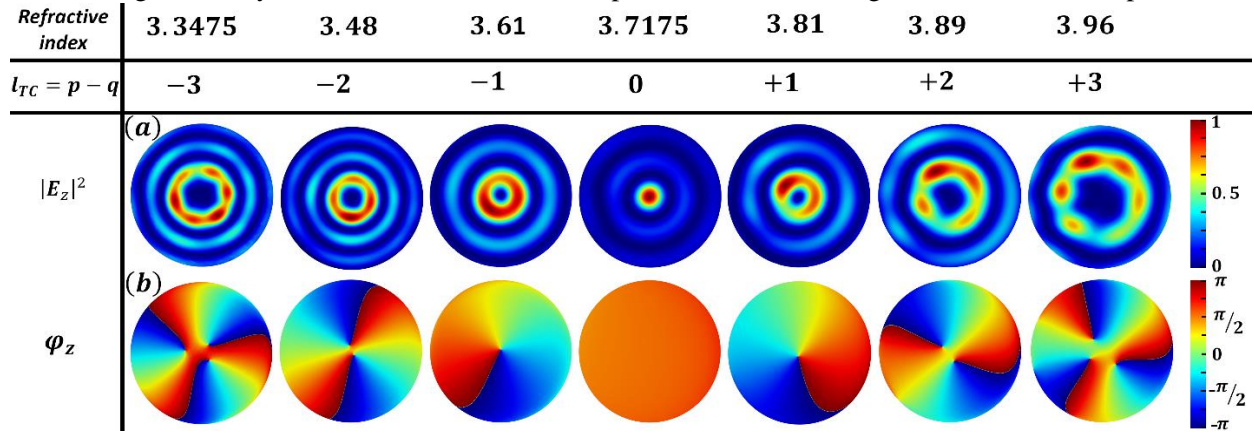

**Figure 7.** Generation of different OAM wave modes at 1550 nm by embedding bilayer 85 nm $Sb_2Se_3$ film rings with an ITO nano-heater between them. Different OAM modes were achieved: (a) Longitudinal component of CVVs ($|E_z|^2$) showing topological charge of OAM waves $l = -3, -2, -1, 0, +1, +2, +3$ at 1550 nm wavelength by considering overall refractive index of $Sb_2Se_3$ as 3.3475, 3.48,3.61, 3.7175,3.81,3.89,3.96 as an effect of partial phase change (b) Corresponding spiral phase pattern for $l = -3, -2, -1, 0, +1, +2, +3$.

$Sb_2Se_3$ layers with an 85 nm ITO nano-heater sandwiched between them compared to design 3 of thicker film.

**Figure 7.** demonstrates the generation of several different OAM modes using design 3 and shows that it is possible to tune seven different OAM modes from $l = -3$ to $l = +3$ by varying the overall effective refractive index from 3.285 (amorphous) to 4.05 (crystalline) using 85 nm bi-layer $Sb_2Se_3$. Here, variation of the overall change of the refractive index is considered as the effect of the phase change of the $Sb_2Se_3$ film from amorphous to crystalline state.

**Conclusion**
In this article, an integrated photonic device for the generation of rapidly tunable OAM waves at 1550 nm using an ultra-low-loss PCM ($Sb_2Se_3$) and an ITO film-based nano-heater has been demonstrated through rigorous numerical modeling. Different design options and their corresponding advantages and drawbacks are described in detail. At first, an FEM analysis is conducted on a silicon ring resonator with grating elements to observe different OAM modes ($l = +3, +2, +1, 0, -1, -2, -3, -4$) at different wavelengths in telecom band and then an ultra-low loss PCM ($Sb_2Se_3$)film with ITO nano-heater is introduce to achieve tunable OAM modes at a fixed wavelength (1550 nm). Initially, a simple design is revealed using $Sb_2Se_3$ film by considering overall change of refractive index of the film due to different volume fractions of the crystallized phase to tune various OAM modes ($l = -3, -2, -1, 0, +1, +2, +3$) at 1550 nm. Then, a critical analysis of different realistically implementable design options with an ITO nano-heater and short pulse voltage to fractionally change the phase of the $Sb_2Se_3$ film is presented. The FEM analysis is conducted to demonstrate the phase change of this design by analyzing the temperature distribution and cooling rate of $Sb_2Se_3$ film to check the potential tunability of the proposed device. A comparative study of ITO nano-heater based reversible phase change of a single layer 105 nm $Sb_2se_3$ with 120 nm ITO ring and a bilayer 85 nm $Sb_2Se_3$ with 85 nm ITO ring on a silicon ring resonator is conducted. According to the thermal analysis, the bilayer thinner $Sb_2Se_3$ film is found to be more suitable over thicker single layer $Sb_2Se_3$ to achieve more OAM modes due to its higher cooling rate providing more possibility reversible switching of the large volume fraction of the PCM film. Finally the OAM generation of $l = -3, -2, -1, 0, +1, +2, +3$ is verified using thinner 85 nm bi-layer $Sb_2Se_3$ film. This proposed design will guide researchers towards implementing ultra-fast, tunable OAM generation or tunable OAM transceivers using on-chip integrated Si photonic devices, thereby opening up an innovative method for secured optical communication with enhanced data rate.

**Funding**

JRH acknowledges support from the Air Force Office of Scientific Research (Program Manager Dr. Gernot Pomrenke) under award number FA9550-20RYCOR059. IA and MSA acknowledge funding through the Small Business Office under award number FA8650-18-P-1700.

**Disclosures**

The authors declare no conflicts of interest.

**Data Availability**

Data underlying the results presented in this paper may be obtained from the authors upon reasonable request.